\documentstyle[11pt]{article}
\newcommand{\beq}{\begin{equation}}
\newcommand{\eeq}{\end{equation}}
\newcommand{\dd}[2]{\frac {\partial #1}{\partial #2}}
\newcommand{\pdr}{\partial}
\newcommand{\beqs}{\begin{eqnarray}}
\newcommand{\eeqs}{\end{eqnarray}}

\newcommand{\eqn}[1]{~(\ref{#1})}

\newcommand{\lb}{\mbox{[}}
\newcommand{\un}[1]{\underline{#1}}
\newcommand{\DOE}{This work was supported in part by U.S. Department of Energy
 Grant No. DE--FG02--91ER40685}
\newcommand{\nn}{\nonumber}
\newcommand{\strf}{f_{\alpha \beta \gamma}}
\newcommand{\half}{\frac{1}{2}}
\newcommand{\eps}{\epsilon}

\newcommand{\I}{\cite{I}}
\newcommand{\pr}{\prime}
\newcommand{\eu}{$\un{E(2)}$}
\newcommand{\myc}{$\un{C_{2}}\;\;$}
\newcommand{\ov}[1]{\frac{1}{#1}}
\newcommand{\fr}[2]{\frac{#1}{#2}}
\newcommand{\al}{\alpha}

\newcommand{\rep}{ representation $\;$}
\newcommand{\cirs}{\stackrel {\scriptstyle \circ}{\scriptstyle \circ}}
\newcommand{\xs}{\stackrel {\scriptscriptstyle \times}
					{\scriptscriptstyle \times} }
\newcommand{\hatt}[1]{\hat{\tilde{#1}}}

\begin{document}

\begin{center}
\vskip 2.5cm
{\LARGE \bf Alternative Canonical Formalism for the Wess-Zumino-Witten
Model}
\vskip 1.0cm
{\Large \dag S.~G.~Rajeev, \dag\dag G.~Sparano and \dag\dag P.~Vitale}
\\
{\large \dag Department of Physics, University of Rochester\\
 Rochester, NY 14623 USA}
\\
{\large  \dag \dag Dipartimento di Scienze Fisiche, Universit\`a di Napoli\\
and I.N.F.N. Sez.\ di Napoli,\\
Mostra d'Oltremare Pad.~19, 80125 Napoli ITALY.}
\vskip 0.4cm
\end{center}

\centerline{\bf Abstract}

We study a canonical  quantization of the Wess--Zumino--Witten (WZW) model
 which depends on two integer parameters rather than one. The usual theory can
 be obtained as a contraction, in which our two parameters go to infinity
 keeping the difference fixed. The quantum theory is equivalent to a
generalized
 Thirring model, with left and right handed fermions transforming
under different representations of the symmetry group. We also point out that
 the  classical WZW model with a compact target space has a canonical formalism
 in which the current algebra is an affine Lie algebra of non--compact type.
 Also, there are some non--unitary quantizations of the WZW model in which
there
 is  invariance only under  half the conformal algebra (one copy of the
 Virasoro algebra).

\vfill
\begin{center}
December 1993
\end{center}
\newpage
\setcounter{footnote}{1}

\newcommand{\tr}{\;{\rm tr}\;}
\section{Introduction}

In \cite{I} one of us discussed a new  canonical
formulation of the
two dimensional nonlinear model. In this paper we will
generalize the formalism to include the Wess--Zumino--Witten
term. The well known \cite{witten},\cite{bhattraj}
quantization of this model
depends on an integer parameter, k; for a particular value of
the
coupling constant the theory is conformally invariant.
 We will find that
our quantization will depend on {\it two} integer
parameters,
the conventional one being the limiting case as they go to
infinity keeping the difference fixed. This more general
quantization scheme
does not lead to any new unitary  conformal field theories; the fixed
point for the $\beta$ function occurs only
in the limiting case. However if we drop the requirement of unitarity,
there are field theories parameterized by a pair of integers, which have
half--conformal invariance (invariance under one copy of the Virasoro
algebra).

The unitary quantum theory is  equivalent to a fermionic  field
theory with four--Fermi interactions; the two integers may be thought of as
the
numbers of the left and right  moving fermions. The fact
that
the conventional quantization of the nonlinear model is the
large $k$ limit of a  fermionic  system was shown in
\cite{polyawieg}. Our work can be viewed as a
generalization of this result. Also, we work within
canonical
quantization rather than path integral quantization as in
Ref.\cite{polyawieg}.
Some work in the same direction has been done in \cite{Bhattacharya}.

Such an equivalence of the WZW model to a fermionic theory
 is  another
example of
the Bose--Fermi correspondence.  Our example is also of interest as
a two--
dimensional analogue of Quantum Chromodynamics (QCD). The
nonlinear model in two dimensions is classically scale
invariant; this is broken at the quantum level. The
coupling
constant is scale dependent and has an ultraviolet stable
zero
at the origin: the theory is asymptotically free.  This
is analogous to the scaling behaviour of QCD.

In more detail, the field equation of the model we study is
\beq
     \pdr^{\mu}(\pdr_{\mu}g\;g^{-1})-
\rho\eps_{\mu\nu}\pdr_{\mu}g g^{-1}\pdr_\nu g g^{-1}=0.
\label{eqmotion}
\eeq
Here, $g:R^{1,1} \longrightarrow G$ is a map from two--dimensional
 Minkowski space to  a simple compact Lie group $G$. In ref. \cite{I} the case
$\rho=0$ was studied. A new canonical  formalism, in which
the current algebra is the direct sum of two affine Lie  algebras, was found.
In
this paper, we will show that this approach can be extended to the case with a
WZW term. The main difference is that the central charges of the two affine Lie
algebras are not equal.

\section{The Standard Canonical   Formalism}

Let us review the classical formulation of the nonlinear
model\cite {witten},\cite{bhattraj}. For a related
discussion see Ref.\cite{I}.

The crucial property of the nonlinear model is that it can
be
formulated
entirely in terms of the current $\pdr_{\mu}gg^{-1}$. The
equation of motion \eqn{eqmotion} is equivalent to the
following pair of
first order
equations,
\beqs
     \dd{I}{t}&=&\dd{J}{x}+\rho[I,J]  \label{eqmo2}  \\
 \dd{J}{t}&=&\dd{I}{x}-[I,J]  \label{bianchi}
\eeqs
The second equation, which is a sort of Bianchi identity, is
the integrability
condition for the existence of a $g:R^{1,1}\longrightarrow G$
satisfying,
 \beqs
  I &=&\dd{g}{t} g^{-1}  \label{defI}  \\
 J &=& \dd{g}{x} g^{-1}  \label{defJ}
\eeqs
If we also impose the boundary condition
\beq
     \lim_{x \rightarrow -\infty}g(x)=1  \label{bcg}
\eeq
the solution  for $g$ is unique. Then equation \eqn{eqmo2}
guarantees that $g$ satisfies the nonlinear model equation
of
motion.
Note that if we had chosen space to be a circle,
\eqn{bianchi}
would not
imply \eqn{defI} and \eqn{defJ}. The solution to these
equations will not
be periodic  in general. If $(I,J)$ is viewed as a
connection,
\eqn{bianchi}
says that it is flat. But in order for a flat connection to
be
`pure gauge'
as in \eqn{defI} and \eqn{defJ}, it is necessary also for
the
parallel transport
operator around a homotopically nontrivial  curve (holonomy)
to be equal
to the identity. It might be interesting to study the theory
on the circle as well.
This first order formulation in
terms of currents
is particularly convenient in two dimensions. In higher
dimensions, the
spatial components
of the currents must satisfy some first class constraints.

In our case, an initial data is given by any  pair of
functions
$I,J:R\longrightarrow \un{G}$ which are square integrable ($\un{G}$ is the Lie
algebra associated to $G$).
The
square
integrability is a condition on how quickly they must decay
to
zero at
infinity. It is needed for finiteness of energy (see
\eqn{ham}
below).

{}From the above discussion, it is clear that $I$ and $J$
provide co--ordinates
on the phase space (space of initial data) of the nonlinear
model. This is
exactly what the $p$ and $q$ variables  do in classical
mechanics. Observables
are functions on the phase space, so they can always be
written as functions
of $I$ and $J$ ( the same way observables of a classical
mechanical system
can be written as functions of $p$ and $q$).
 Just as  the Canonical Commutation Relations (CCR)
determine the Poisson brackets of any
two observables in classical mechanics, the commutation
relations
of $I$ and $J$
determine completely the algebra of observables.
Like the CCR,
 the Poisson brackets
of $I$ and $J$ also define a Lie algebra (the  current
algebra).

The standard action  for the WZW model  is
\beq
     S={1\over 4\lambda^2}\int \tr \pdr_\mu g \pdr^\mu g^{-
1}
d^2x+n\Gamma.
\eeq
Here the WZW term  $\Gamma$,
\beq
     \Gamma= {1\over 24\pi}\int_B  d^3y \eps^{ijk}\tr
\pdr_igg^{-1}\pdr_jgg^{-1}\pdr_k g g^{-1}\label{Gamma}
\eeq
is an integral over a three manifold $B$ with space--time
as
boundary. We recover the  above equation of motion  with
\beq
     \rho={n\lambda^2\over 4\pi}.
\eeq
Note that the action as well as the canonical formalism
depends on  the extra parameter $\lambda$, which does not
affect the  classical dynamics as it cancels out. In the
quantum theory $n$ must be an integer so that $e^{iS}$ be independent
of the extension to the third dimension.

It is possible to derive a set of Poisson Brackets for $I$
and
$J$ from the action principle.
 But for our purposes, it is better to regard the Poisson
brackets and a hamiltonian as the basic postulates,
justified
by reproducing the equation of motion.  This will allow us
to
generalize to a situation where the action is not obvious.

There is an extensive literature on the philosophy
of current algebras, see \cite{god}  for more
references.
This approach  had mixed success in theory of strong
interactions. For the two
dimensional nonlinear model on the other hand, this seems to
be the most
natural point of view.

In our case the Poisson brackets of $I$  and $J$ are,
\beqs
 \frac{1}{2 \lambda^{2}} \{ I_{\alpha}(x) ,I_{\beta}(y)
\}_{1}
&=&
     f_{\alpha \beta \gamma} I_{\gamma}(x)\delta (x-y) +\rho
f_{\alpha\beta\gamma}J_\gamma\delta(x-y)
\label{PBII}  \\
 \frac{1}{2 \lambda^{2}}\{ I_{\alpha}(x) ,J_{\beta}(y)
\}_{1}
&=&
  f_{\alpha \beta \gamma} J_{\gamma}(x)
     \delta (x-y) -\delta_{\alpha \beta} \delta^{\prime}(x-
y)
\label{PBIJ}\\
 \frac{1}{2 \lambda^{2}}\{ J_{\alpha}(x) ,J_{\beta}(y)
\}_{1}
&=& 0.\label{PBJJ}
\eeqs
The subscript on the brackets distinguishes these ones from
other
Poisson brackets
 we  will introduce later.

These brackets can be written more elegantly  if we
introduce
some mathematical notation.
Let us define the infinite dimensional Lie group $RG$ to be
the set of
functions $g$ satisfying \eqn{bcg}:
\beq
       RG=\{g:R\longrightarrow G\;\mid \; g(-\infty)=1\}
\eeq
The multiplication in $RG$ is just pointwise:
\beq
     g_{1}g_{2}(x)=g_{1}(x)g_{2}(x)
\eeq
As always, the Lie algebra of $RG$ will be denoted by an
underline:
\beq
     \un{RG}=\{\xi:R \longrightarrow \un{G}
               \:\mid\:\int\;tr\;\xi^{2}\;dx<\infty\}
\eeq
The  Lie bracket in $\un{{RG}}$ is also pointwise.
Occasionally, we will need
another  Lie algebra with the {\em same} vector space as
above, but for which all brackets
are zero. This abelian Lie algebra will be called
$\un{A}$. It
is important
to distinguish $\un{RG}$ from $\un{A}$ although  as vector
spaces, they are  the same  function space on $R$.

Since the Lie algebra $\un{G}$ has a natural invariant inner
product,
we will identify it with its dual as a vector space.

Define now the Lie algebra  $\un{C_{1}}$  to be the set of
triples $(\xi,\eta,b)$ ,where $\xi$ and $\eta$
are square
integrable functions from $R$ to $\un{G}$, $b$ is a real
number. The Lie bracket
is defined to be
\beqs
\lefteqn{ \lb (\xi,\eta,b)\;\;
,\;\;(\xi^{\prime},\eta^{\prime},b^{\prime})]=} \nn \\
  & &([\xi,\xi^{\prime}], [\xi,\eta^{\prime}]-
[\xi^{\prime},\eta^{\prime}]+\rho[\xi,\xi'],
-\int\;tr\;(\xi\dd{}{x}\eta^{\prime}-
\xi^{\prime}\dd{}{x}\eta)\;dx)
\eeqs

This algebra has  an abelian subalgebra isomorphic to
$\un{A}\oplus R$ when
$\xi=0$. If $\rho=0$,  $\un{C_{1}}$ is the semi--direct sum
of
two subalgebras.
\beq
     \un{C_{1}}=\un{RG}\dot{\oplus}(\un{A}\oplus R)
\eeq
where the dot denotes a semi-direct sum of Lie algebras. By
a
change of basis, we can see that if $\rho\neq 0$, this is a
semi--direct sum of an affine Lie algebra with an abelian
algebra. Then $\un{C_1}$ describes the algebra of currents given by
\eqn{PBII}, \eqn{PBIJ}, \eqn{PBJJ}, for any value of $\rho$.

The hamiltonian describing the non linear model is
\beq
     H_{1}=\frac{1}{4\lambda^{2}}\int tr(I^2 +J^2) dx.
\label{ham}
\eeq
This can of course  be derived from the usual action
principle; but we prefer
to justify it directly  by showing that, together with \eqn{PBJJ}, it leads to
the required equations
of motion \eqn{eqmo2} and \eqn{bianchi}. This is a
straightforward calculation:
\beqs
 \dd{I_{\beta}(y)}{t}&=& \{ H_{1},I_{\beta}(y) \}_{1}
=\dd{J_{\beta}(y)}{x} +\rho f_{\alpha\beta\gamma} I_\alpha(y)
J_\gamma(y) \\
 \dd{J_{\beta}(y)}{t}&=& \{ H_{1},J_{\beta}(y) \}_{1}=
\dd{I_{\beta}(y)}{x}
               +f_{\alpha
\beta \gamma}I_{\alpha}(y)J_{\gamma}(y)
{}.
\eeqs
Note that the parameter $\lambda$ ( the `coupling constant')
drops out of the equations of motion.

For the record, let us note the stress tensor
$\Theta_{\mu\nu}$ of the theory
as a function of the currents.
\beqs
 \Theta_{00}=\Theta_{11}&=&\frac{1}{4\lambda^{2}}tr(I^2
+J^2)
\\
   \Theta_{01}=\Theta_{10}&=&\frac{1}{2\lambda^{2}} tr(IJ)
\eeqs
This is traceless and conserved, so this formalism is
conformally and Poincar\'e
invariant. We  won't discuss the Poisson brackets of
$\Theta_{\mu\nu}$, since
they will be derived in a more general context later.

\section{The contraction of \un{SU(2)} to \un{E(2)}}

There is a certain analogy between our current algebra and the Euclidean group.
The spatial currents $J$ commute, as translations do, while the time components
$I$ are analogous to rotations. One way to construct a representation of the
Euclidean group is to diagonalize the commuting generators, and then
representing rotations as differential operators in terms of them (the method
of induced representations).

 The analogous procedure in our case would be to diagonalize $J$, which is
essentially the Schr${\ddot o}$dinger representation of the field theory.
Unfortunately
this is very complicated to do in practice, since an appropriate measure of
integration in the configuration space does not exist. The only known measure
is
an analogue of the
Wiener measure \cite{frankrep} which unfortunately is not the
appropriate one. Even in  the case of  the abelian group (where the required
measure is  known and is a Gaussian), the
inner product  which makes the hamiltonian self--adjoint is not
 obtained from the Wiener measure. So we look at another way
of constructing representations of algebras with abelian
ideals, the method of contractions.

We will  describe  here the  construction of  the
representation of
$\un{E(2)}$, using the method
of `contraction'
or `deformation' of Lie algebras.
 The discussion follows an article of Inonu \cite{inonu}.
This will suggest another  way of dealing
with the nonlinear model.

Recall that  $\un{E(2)}$ is the Lie algebra of  isometries
of
$R^{2}$. It
is a non--compact Lie algebra, this implying that its unitary
representations
are infinite dimensional.

Now geometrically, we can regard $R^{2}$ to be  the limit
as
the radius
goes to infinity of a two dimensional sphere, $S^{2}$. There
must therefore
be some way of thinking of \eu~  as a limit of the isometry
algebra of
$S^{2}$. This  is what the method of contractions (or,
deformations)
of Lie algebras does. Another  example is the Galilei group
which is the
limit as the velocity of light goes to infinity of the
Lorentz
group. The
 Poincar\'e group is a contraction of the group $O(2,3)$.

In all these examples a non--simple group is obtained as the
contraction
of a simple group. This is useful in representation theory
because the
representations of these simple groups are in many ways
easier
to understand.
 We will work out the case of \eu~  in detail here.

The isometries of $S^{2}$ form the Lie algebra $\un{SU(2)}$.
The generators  of $\un{SU(2)}$ are the three rotations
$S_{i}$, satisfying
\beq
     \lb S_{i},S_{j}]=\eps_{ijk}S_{k}
\eeq
Let us now introduce a real  parameter $\tau$ and define
\beqs
     S&=&S_{3}\nn \\
     P_{a}&=&\tau S_{a}\;  ; a=1,2
\eeqs
Then, the commutation relations become
\beqs
     \lb S,P_{a}]_{2}&=&\epsilon_{ab}P_{b}  \nn \\
     \lb P_{a},P_{b}]_{2}&=&\tau^{2}\eps_{ab}S \label{su2}
\eeqs
the subscript being present to distinguish it from the
relations of \eu.
For any finite value of $\tau$ the above relations describe
 $\un{SU(2)}$ in a peculiar choice of basis. But in the
limit
as
 $\tau\rightarrow 0$ they become just \eu.

The irreducible representations  of $\un{SU(2)}$ are labelled by
spin
$j$,
 which is either integer or half integer. The dimension of
the
spin $j$
representation is $2j+1$. If a representation of
$\un{SU(2)}$
is to tend
to one for \eu~  in the limit as $\tau\rightarrow 0$, $j$ must
go to infinity
at the same time. For, all unitary representations of \eu~
are infinite dimensional.

Let us now do this explicitly. An orthonormal
 basis $|m>$ for the spin $j$ representation of
$\un{SU(2)}$ is labelled by $m=-j,-j+1.... j-1,j$. The matrix
elements are
\cite{inonu}
\beqs
<m|\hat{S}|n> &=& ~~m \delta_{m,n} \nn \\
<m|\hat{S}_{1}|n> &=& ~~\half\surd(j-m)(j+n)\delta_{n,m+1}\nn \\
&  &   +\half\surd(j+m)(j-n)\delta_{n,m-1}\nn \\
   <m|\hat{S}_{2}|n>&=&-\frac{i}{2}\surd(j-
m)(j+n)\delta_{n,m+1}\nn \\
               & &+\frac{i}{2}\surd(j+m)(j-n)\delta_{n,m-1}\nn
\eeqs

As we already mentioned, to have a meaningful limit as  $\tau\rightarrow 0$,
we must
also let $j\rightarrow \infty$ such that
\beq
       \lim \tau j=a \nn
\eeq
is finite. In this limit the matrix elements of $S$ and $P$
are,
\beqs
          <m|\hat{S}|n>&=&m\delta_{m,n} \nn \\
 <m|P_{1}|n>&=&\frac{a}{2}(\delta_{n,m-1}+\delta_{n,m+1}) \nn \\
  <m|P_{2}|n>&=&\frac{ia}{2}(\delta_{n,m-1}-
\delta_{n,m+1}) \nn
 \eeqs
 which is what we want.

\section{New   Canonical  Formalism}

We will now show that the current algebra $\un{C_1}$ is the
contraction of the direct sum of two affine Lie algebras. Thus
we can construct representations of it as limiting cases of
representations of affine Lie algebras. However, it turns out
that there is a more general canonical formalism for the WZW
model where the current algebra is just a pair of affine Lie
algebras. This does not follow from the  usual action
principle (an action  principle which leads to this general
canonical formalism can be found).
It is also
found that this new formalism preserves conformal (and hence
Poincar\'e) invariance
at the classical level.

 Furthermore,  the new  current algebra  is isomorphic
to
that of
fermion currents. So a representation can be found in terms
of
fermion
bilinears. Then the Hamiltonian of the WZW
model
becomes identical
to that of a  Thirring model. This is a bosonization of the
Thirring model. The special case without WZW term was
discussed in \I.

By analogy to  \eqn{su2} (and to $\un{E(3)}$ as discussed in
\I) we are
lead to a deformation of the Lie algebra
$\un{C_{1}}$. We will attempt to deform the algebra of $I$ and
$J$ such that the equations of motion remain unchanged. Let us
make the ansatz ( $a,\eps,\tau,\mu$ are parameters which we
assume to be real for now):
\beqs
 \frac{1}{2 \lambda^{2}} \{ I_{\alpha}(x) ,I_{\beta}(y)
\}_{1\half} &=&
     f_{\alpha \beta \gamma} I_{\gamma}(x)\delta (x-y)\nn \\
& & + af_{\alpha \beta \gamma}J_\gamma (x)\delta(x-y)
\label{PBII1}  \\
 \frac{1}{2 \lambda^{2}}\{ I_{\alpha}(x) ,J_{\beta}(y)
\}_{1\half} &=&
  f_{\alpha \beta \gamma} J_{\gamma}(x)
     \delta (x-y) -\delta_{\alpha \beta} \delta^{\prime}(x-
y)\nn \\
& &+ \eps f_{\alpha \beta \gamma} I_\gamma (x)\delta(x-y)
\label{PBIJ1}\\
 \frac{1}{2 \lambda^{2}}\{ J_{\alpha}(x) ,J_{\beta}(y)
\}_{1\half} &=&
     \tau^{2}\strf I_{\gamma}(x)\delta(x-y) \nn \\
& &+ \mu f_{\alpha\beta
\gamma} J_\gamma(x)\delta(x-y)  \label{PBJJ1}
\eeqs
together with the hamiltonian \eqn{ham} these Poisson brackets  lead to the
equations,
 \beqs
     \dd{I}{t}&=&\dd{J}{x}+(a-\eps)[I,J]    \\
 \dd{J}{t}&=&\dd{I}{x}+(1-\tau^{2})[I,J]
\eeqs
There is an unwanted factor of $(1-\tau^{2})$, so we don't
 quite get the
equations
\eqn{eqmo2},\eqn{bianchi} we want.
But if we rescale $I$ and $J$ by $(1-\tau^{2})$ this problem
disappears. The equations become
 \beqs
     \dd{I}{t}&=&\dd{J}{x}+{(a-\eps)\over (1-\tau^2)}[I,J]
\\
 \dd{J}{t}&=&\dd{I}{x}+[I,J]
\eeqs
which agree with \eqn{eqmo2},\eqn{bianchi} if
\beq
     \rho={a-\eps\over 1-\tau^2}.
\eeq

Thus we arrive at the algebra
\newpage
\beqs
 \frac{1}{2 \lambda^{2}} \{ I_{\alpha}(x) ,I_{\beta}(y)
\}_{1\half} &=&
     (1-\tau^2)f_{\alpha \beta \gamma} I_{\gamma}(x)\delta (x-
y)\cr
& &+a(1-\tau^2)f_{\alpha \beta \gamma}J_\gamma (x)\delta(x-y)
\label{PBII2}  \\
 \frac{1}{2 \lambda^{2}}\{ I_{\alpha}(x) ,J_{\beta}(y)
\}_{1\half} &=&
  (1-\tau^2)f_{\alpha \beta \gamma} J_{\gamma}(x)
     \delta (x-y) \nn \\
& & -(1-\tau^2)^2\delta_{\alpha \beta}
\delta^{\prime}(x-y)\cr
& &+ (1-\tau^2)\eps f_{\alpha \beta \gamma} I_\gamma
(x)\delta(x-y)
\label{PBIJ2}\\
 \frac{1}{2 \lambda^{2}}\{ J_{\alpha}(x) ,J_{\beta}(y)
\}_{1\half} &=&
     \tau^{2}(1-\tau^2)^2\strf I_{\gamma}(x)\delta(x-y)\cr
& &+(1-\tau^2)\mu f_{\alpha\beta\gamma} J_\gamma(x)\delta(x-y).
\label{PBJJ2}
\eeqs
Let us call this algebra $\un{C_2}$.
This is supplemented with the rescaled hamiltonian,
\beq
  H_{2}=\frac{1}{4\lambda^{2}(1-\tau^{2})^{2}}\int tr(I^2
+J^2) dx.  \label{ham2}
\eeq

\noindent Clearly in the limit $\tau\rightarrow 0$ we recover the
standard formalism. We have just shown that this formalism
leads to our
equations of motion,  with the above identification of $\rho$.

The Poisson bracket relations above are an obscure way of
writing
$\un{C_{2}}$. Without
a tedious calculation it is not even clear that they obey the
Jacobi identities. But there is a change of variables which
simplifies them
substantially. It is suggested by the analogy of $\un{C_{1}}$
to
$\un{E(3)}$ and
$\un{C_{2}}$ to $\un{SU(2)}\oplus\un{SU(2)}$ \cite{I}.

Let us define $L$ and $R$ by
\beqs
     I&=&2\lambda^{2}(1-\tau^{2})(\alpha L+\beta R)   \nn \\
     J&=&2\tau\lambda^{2}(1-\tau^{2})(\gamma L+\delta R) .
\label{defR}
\eeqs
with $L$ and $R$ generators of two commuting affine Lie algebras,
\beqs
\{ L_{\alpha}(x) ,L_{\beta}(y) \}_{2} &=&
  f_{\alpha \beta \gamma} L_{\gamma}(x)
     \delta (x-y) +\frac{k}{2\pi}\delta_{\alpha \beta}
\delta^{\prime}(x-y)  \label{LL}\\
\{ R_{\alpha}(x) ,R_{\beta}(y) \}_{2} &=&
  f_{\alpha \beta \gamma} R_{\gamma}(x)
     \delta (x-y) -\frac{\bar k}{2\pi}\delta_{\alpha \beta}
\delta^{\prime}(x-y)  \label{RR}\\
\{ L_{\alpha}(x) ,R_{\beta}(y) \}_{2} &=&0
\label{LR}
\eeqs
and $k,\bar k $ are a pair of constants.

It is now straightforward (although quite tedious) to show
that this algebra goes over to $\un{C_2}$ under the above
change of variables, if we choose
\beqs
\alpha=1-\rho\tau\;\;& &
                \beta=1+\rho\tau\\
\gamma=\tau(\rho\tau-1)\;\;& &\delta=\tau(\rho\tau+1)\\
\eps=\mu=\rho\tau^2\;\;& & a=\rho \\
k={\pi\over 2\lambda^2\tau(1-\rho\tau)^2}\;\; & &
\bar k={\pi\over 2\lambda^2\tau(1+\rho\tau)^2}
\eeqs
Thus our current algebra $\un{C_2}$ is isomorphic to a direct
sum of two affine Lie algebras.
 These Poisson brackets
can be  derived from
an action principle as in ref.\I, but we will postpone that
discussion.

 Note that the change of variables to $L$ and $R$ is
singular
if $\tau=0$.
The equations \eqn{PBII2}-\eqn{PBJJ2} are still well defined
in this limit, it is just
that at $\tau=0$ $\un{C_2}$ is no longer isomorphic to the sum of two affine
Lie
algebras since
$L$ and $R$ don't exist.  On the other hand at $\tau=\pm 1$,
$L$ and $R$
Poisson brackets
exist, but not those of $I$ and $J$.

It is more convenient to use $L$  and $R$  as the basic dynamical
variables, since they have simple Poisson brackets.  The
hamiltonian in this
language is
\beq
     H_2= \lambda^2(1+\tau^2)\int \tr[ (1-\rho\tau)^2L^2+2{1-
\tau^2\over
1+\tau^2}(1-\rho^2\tau^2)LR+(1+\rho\tau)^2R^2] dx\label{2ham}
\eeq

Also, a  choice of independent parameters is $\rho,k,\bar k$.
We
find for instance that
\beq
     {2\pi  \rho\over \lambda^2}=(k-\bar k)
\frac {16 k \bar k} {(\surd k+\surd {\bar k})^4}.
\eeq

Thus we see that $k$ and $\bar k$ must have the  same sign for $\rho$ to be
real. Without loss of generality we can choose
them to be positive. The limit $\tau \to 0$ which should be
the conventional
theory corresponds to letting $k,~\bar k \to \infty$ keeping $k-
\bar k$ fixed and equal to an even integer. In this limit,
\beq
     {2\pi \rho\over \lambda^2}=k-\bar k
\eeq
so that we can identify $n={k-\bar k\over 2}$.  In general it
is not necessary
that this quantity be an integer, it can take any value
determined as above
by a pair of positive integers.

Clearly this general canonical quantization cannot be obtained
by the standard
action principle. We can  now find ( following \cite{I}) an
action principle which gives this canonical formalism. We define variables
$l,r:R^{1,1}\to G$
such that
\beq
     L={k\over 2\pi} {\pdr l\over \pdr x} l^{-1},\;\; R={\bar
k\over 2\pi} {\pdr
r\over \pdr x} r^{-1}
\eeq
Now define the action
\beqs
     S_2&=&k\Gamma(l)-\bar k\Gamma(r)
 -\lambda^2(1+\tau^2)\int \tr[ (1-\rho\tau)^2L^2\nn \\
& &+2{1-\tau^2\over
1+\tau^2}(1-\rho^2\tau^2)LR+(1+\rho\tau)^2R^2] dx dt
\eeqs
where $\Gamma$ is the  WZW term defined in \eqn{Gamma}.
Arguments exactly
analogous to those in Ref.\cite{I} show that this leads to the
new canonical
formalism. For $e^{iS}$ to be single valued, $k$ and
$\bar k$ must be
integers. We will obtain the same requirement from the
representation theory of
the affine Lie algebra.
This form of the action does not look
Lorentz
invariant, but we will show that the theory is in fact
classically conformal
invariant. This implies in particular Lorentz invariance.

The components of the stress tensor are,
\beqs
     \Theta_{00}=\Theta_{11}&=& {1\over 4\lambda^2(1-
\tau^2)^2}\tr[I^2+J^2]\\
\Theta_{01}=\Theta_{10}&=& {1\over 2\lambda^2(1-
\tau^2)^2}\tr[IJ]
\eeqs
It is often more convenient to use instead the quantities
\beq
     \Theta=-\half(\Theta_{00}+\Theta_{01})\quad
\tilde \Theta=\half(\Theta_{00}-\Theta_{01}).
\eeq
Classical conformal invariance amounts to the statements
\beq
     (\pdr_t-\pdr_x)\Theta=0\quad (\pdr_t+\pdr_x)\tilde
\Theta=0
\eeq
together with the the Poisson brackets
\beq
     \{\Theta(u),\Theta(v) \}=\Theta([u,v])
\eeq
\beq
\{\tilde  \Theta(u),\tilde  \Theta(v)\}=\tilde \Theta([u,v])
\eeq
\beq
\{ \Theta(u),\tilde  \Theta(v)\}=0.
\eeq
Here $\Theta(u)=\int dx u(x)\Theta(x)$ etc.

In general, if $M_A$ are currents satisfying an affine Lie
algebra (in the sense of Poisson brackets),
\beq
     \{M_A(x),M_B(y)\}=f^C_{AB}M_C(x)\delta(x-
y)+{\Omega_{AB}\over 2\pi}\delta'(x-y)
\eeq
the conditions for
\beq
\Theta(x)=G^{AB}M_A(x)M_B(x)
\eeq
\beq
\tilde \Theta(x)=\tilde G^{AB}M_A(x)M_B(x)
\eeq
to satisfy the previous
Poisson brackets are
\beq
     G^{AB}=2G^{AC}{\Omega_{CD}\over 2\pi}G^{DB}
\eeq
\beq
     \tilde G^{AB}=2\tilde G^{AC}{\Omega_{CD}\over 2\pi}G^{DB}
\eeq
\beq
     0 =2\tilde G^{AC}{\Omega_{CD}\over 2\pi}G^{DB}.
\eeq
These relations may be regarded as the classical analogue of the Master
Virasoro equation
\cite{halpern}. In our case, we can choose $M=(L,R)$, and
write
$\Theta,\tilde \Theta$ as quadratic expressions in $L$ and $R$,

\beqs
     \Theta&=&-{1\over 8\lambda^2(1-\tau^2)^2}(I+J)^2=
G^{AB}M_AM_B \label{ts1} \\
     \tilde \Theta&=&-{1\over 8\lambda^2(1-\tau^2)^2}(I-
J)^2=\tilde  G^{AB}M_AM_B. \label{ts2}
\eeqs

 The matrices $G^{AB}$ and $\tilde  G^{AB}$ are
\beq
     G^{AB}=-\frac{\lambda^2}{2}
\left(
\begin{array}{cc}

(1-\rho\tau)^2(1-\tau)^2 {\bf 1}_3&
                              (1-\rho^2\tau^2)(1-\tau^2){\bf 1}_3 \\
(1-\rho^2\tau^2)(1-\tau^2){\bf 1}_3&
                           (1+\rho\tau)^2(1+\tau)^2 {\bf 1}_3
\end{array}
\right)
\eeq

and

\beq
\tilde  G^{AB}= ~{\lambda^2\over 2}
\left(
\begin{array}{cc}
(1-\rho\tau)^2(1+\tau)^2 {\bf 1}_3&
                              (1-\rho^2\tau^2)(1-\tau^2){\bf 1}_3\\
(1-\rho^2\tau^2)(1-\tau^2){\bf 1}_3&
                           (1+\rho\tau)^2(1-\tau)^2 {\bf 1}_3
\end{array}
\right)
\eeq
and the central term is given by
\beq
     \Omega^{AB}=
\left(
\begin{array}{cc}
k{\bf 1}_3&0\\ 0&-\bar k {\bf 1}_3
\end{array}
\right)
\eeq
where by ${\bf 1}_3$ we mean the $3 \times 3$ identity matrix.
It is now straightforward to check that the conditions above
for classical conformal invariance are in fact satisfied.
The conditions for conformal invariance at the quantum level
are more stringent; we will not find any new unitary conformal
field theories in our approach.

\section{Quantization}

So far we have mainly talked about an alternative classical
formulation of
the nonlinear
model. Now let us begin quantizing the new formalism.

Recall that $\un{C_{2}}$ is the analogue of the CCR in the
nonlinear
model. Therefore finding a representation of \myc is the first
step in
quantizing the nonlinear model.
 We need a unitary representation to construct a quantum
theory. It is
desirable
also that the representation be irreducible. For, each
subrepresentation would
otherwise form an independent dynamical system (this is why
one always
picks an irreducible representation for the CCR in ordinary
quantization).  Once operators representing $L$ and $R$
are found, we can try to construct the hamiltonian and other
observables
as their functions. But this last step is going to involve a
renormalization.

Representations of our current algebra can be classified  by
purely algebraic methods. The earlier considerations require
that $k$ and $\bar k$ be both positive. Since $k,\bar k$
appear with opposite signs in the central terms of $L$ and
$R$, we find that the representations of $L$ must be highest
weight and that of $R$ lowest weight, if the representation is
to be unitary.

For $k=1$, there is a unique representation for the current algebra, which can
be written in terms of fermions \cite{witten}.
 For higher level numbers, fermionic representations are a direct
sum of several irreducible representations. The multiplicity
of a given
irreducible representation is finite if a condition, known as
the
Goddard--Nahm--Olive condition (GNO), \cite{god,gno}, is satisfied. Then,
at least
when the multiplicity is
finite, the nonlinear model is equivalent to one sector of
the
fermionic theory. Conversely, the fermionic theory breaks up
into a
set of nonlinear models which don't interact with each other.

It is convenient to use a countable basis for the Lie algebra
when discussing representations. It is possible to
introduce such a basis on the
space of functions on the real line, if we consider only
functions which vanish at  infinity (to have finite energy, our
variables $L,R$ must  in fact
vanish at infinity).
A convenient basis on the space of functions on $R$ vanishing
at infinity is,
\beq
     e_m^{r}(x)=\left({1\over \pi[1+x^2]}\right)^{r}\left({i-
x\over i+x}\right)^m.
\eeq

\noindent We should use $r=1$  for currents and $r=\half$ for spinor
fields.
This is because  $e_m^{\half}$ are orthonormal
(so that Canonical Anti--commutation Relations simplify),
while $e_m^{1}$ satisfy

\beqs
     \sum_{m\in Z}e_{m}^{1}(x)e_{p-m}^{1}(y) &=& e_{p}^{1}(x-
y)\delta(x-y) \\
       \sum_{m\in Z}ime_{m}^{1}(x)e_{-m}^{1}(y) &=&
               \ov{2\pi}\delta^{\pr}(x-y).  \label{e1}
\eeqs

With
\beq
     L_\alpha(x)=\sum_m L_{m\alpha}e^1_m(x)~~~~
 R_\alpha(x)=\sum_m R_{m\alpha}e^1_m(x)
\eeq
we have the  relations

\beqs
     \{ L_{m\alpha},L_{n\beta}\}&=&\strf L_{m+n\;\gamma}
          -i\;k\;m\delta_{\alpha\beta}\delta_{m,-n}\\
          \{ R_{m\alpha},R_{n\beta}\}&=&\strf R_{m+n\;\gamma}
          +i\;\bar k\;m\delta_{\alpha\beta}\delta_{m,-n}.
\eeqs

A unitary representation of this algebra will be given by a
set of
 operators satisfying
\beqs
     L_{m\al}^{\dag}=L_{-m\al}& & R_{m\al}^{\dag}=R_{-m\al}\;\;{\rm ( unitarity
)}\\
     \lb \hat{L}_{m\al},\hat{L}_{n\beta}]&=&
       i\;\strf\;\hat{L}_{m+n\;\gamma}+
               km\delta_{\al\beta}\delta_{m,-n}\\
\lb \hat{R}_{m\al},\hat{R}_{n\beta}]&=&
       i\;\strf\;\hat{R}_{m+n\;\gamma}-
               \bar{k}m\delta_{\al\beta}\delta_{m,-n}
\eeqs
The representation of $L$ can be chosen to be highest weight
(i.e., there exists a vector with $L_m|0>=0$ for $m>0$) and
unitary.
 In this case $k$ is required to be a positive  integer.
 Then the representation
for $R$ has to be lowest weight if we want it to be unitary as well,
since $k,\bar k$ appear with
opposite signs in the central terms
 (recall that $k,\bar k$ must have the same sign from our
earlier
arguments).
A lowest weight representation is just the
complex conjugate of a highest weight representation.
 Hence, {\it  to every pair of irreducible highest weight
representations  of the affine Lie algebra
of $G$, there is a quantization of the WZW model.}

Thus we have a state $|0>$ from which all other states  in the
Hilbert space
 can be constructed, satisfying
\beqs
       \hat{L}_{m\al}|0>=0 \;\;\mbox{ ,$m>0$} \nn \\
     \hat{L}_{0\alpha}|0>=0 \;\;  \mbox{,$\al>0$}\nn \\
       \hat{R}_{m\al}|0>=0 \;\;\mbox{, $m<0$}\nn \\
     \hat{R}_{0\alpha}|0>=0 \;\;  \mbox{,$\al<0$} \nn \\
     \hat{L}_{0\al}|0>=\al.\mu|0>  \;\; \al\in\un{H}\nn \\
     \hat{R}_{0\al}|0>=-\al.\bar \mu|0>  \;\;\al\in\un{H} \nn
\eeqs
This is the statement that $|0>$ is a highest weight vector
for $\hat{L}$
and a lowest weight vector for $\hat{R}$. Here, $\al>0$ means that
$\al$ is a positive
root of $\un{G}$ .Also, $\un{H}$ is the Cartan subalgebra of $\un{G}$
and $\mu,\bar \mu $
the highest weights of some  representations. The necessary and sufficient
condition for
such a \rep to exist is \cite{god,kac}  that $k,\bar k $ be
integers with
\beq
     k\geq\psi.\mu\geq 0,\quad   \bar k\geq\psi.\bar \mu\geq 0
\eeq
where $\psi$ is  the highest root of $\un{G}$ (in general $\psi$ is
normalized to have
length $\surd 2$).

The vector space for the representation labelled by $(k,\mu)$
is constructed
by acting on $|0>$  by the `raising operators'
$\{\hat{L}_{m\al},\;m<0\;$and
$\,m=0,\al<0\}$ and
$\{\hat{R}_{m\al},\;m>0\;$and$\,m=0,\al>0\}$. This is, however,
a  standard construction described for example in \cite {kac,god}. It should
be kept in mind that the
state $|0>$ is  not in general the    ground state. The true ground state
 of the theory is
the one that
minimizes $H$ and is in general quite different from $|0>$.

Let us define the normal ordering
 (suspending the summation convention on $\al$),
\beqs
  \xs \hat{L}_{m\al}\hat{L}_{n\al}
\xs&=&\hat{L}_{n\al}\hat{R}_{m\al}
     ,\;\;n<0\nn \\
&=&\hat{L}_{m\al}\hat{L}_{n\beta},\,\,n\geq 0
                         \label{xs}\\
  \xs \hat{R}_{m\al}\hat{R}_{n\beta}
\xs&=&\hat{R}_{n\beta}\hat{R}_{m\al},
     \;\; n>0\nn \\
                        &=&\hat{R}_{m\al}\hat{R}_{n\beta},\,
, n\leq 0.
\eeqs

Then a representation of the direct sum of two Virasoro algebras, say $\un
{V_1} \oplus {V_2}$, may be
constructed as
\beqs
     \hat{T}(x)&=& \frac{\pi}{k+\tilde {h}(\un{G})}
               \xs \hat{L}_{\al}(x)\hat{L}_{\al}(x) \xs \\
     \hatt{T}(x)&=& -\frac{\pi}{\bar k+\tilde {h}(\un{G})}
               \xs \hat{R}_{\al}(x)\hat{R}_{\al}(x) \xs.
\label{hatt}
\eeqs
Note that the factor in front of $L^{2}$ and $R^2$
has changed from the classical value.
 $k,\bar k$ have been shifted
by an integer $\tilde {h}(\un{G})$ characteristic of the
algebra, called
the `dual Coxeter number' \cite{god}:
\beq
         \tilde {h}(\un{G})\delta_{\al\eps}=\half \strf
f_{\eps\beta\gamma}.
\eeq
These will then satisfy
\beqs
 \lb\ov{i}\int\;u(y)\hat{T}(y)\;dy,\ov{i}\hat{L}_{\al}(x)]&=&
                         u(x)\ov{i}\dd{\hat{L}_{\al}}{x}\\
 \lb\ov{i}\int\;u(y)\hatt{T}(y)\;dy,\ov{i}\hat{R}_{\al}(x)]&=&
                         u(x)\ov{i}\dd{\hat{R}_{\al}}{x}
\eeqs
and
\beqs
 \lb\ov{i}\hat{T}(u),\ov{i}\hat{T}(v)]&=&\ov{i}\hat{T}([u,v])+
                         c\;\omega(u,v)\\
\lb\ov{i}\hatt{T}(u),\ov{i}\hatt{T}(v)]&=&\ov{i}\hatt{T}([u,v])+
                         \bar c\;\omega(u,v).
\eeqs

It is the presence of the extra central term on the  right hand side of these
relations which makes
$\hat{T}$, $\hat{\tilde T}$  satisfying $\un {V_1} \oplus {V_2}$
rather than the algebra of vector fields on the real
line, as in the
classical theory.
  The cocycle $\omega$ is
\beq
 \omega(u,v)=\fr{\pi i}{12}\int
   [u\biggl(\frac{d^{3}}{dx^{3}}+\frac{d}{dx}\biggr)v
                         - u\leftrightarrow v]\;dx
 \eeq
The central charges are
\beq
     c=\frac{k\;\mbox{dim}\;\un{G}}{k+\tilde {h}(\un{G})}\quad
 \bar c=\frac{\bar k\;\mbox{dim}\;\un{G}}{\bar
k+\tilde {h}(\un{G})}.
\eeq

The quantities $T,\tilde  T$ do not however define components
of the true
(physical) stress tensor; these are given by $\Theta$ and
$\tilde  \Theta$.
In general the physical stress tensor does not satisfy $\un
{V_1} \oplus {V_2}$, so that the theory is not conformally invariant.
 $\hat T,\hat {\tilde  T}$ are still useful to construct the
momentum density,
\beq
     \hat \Theta_{01}(x)= \hat T(x)+\hat{\tilde  T}(x).
\eeq
The hamiltonian  is more involved:
\beq
     H=\int \Theta_{00}(x)dx=\int [{1+\tau^2\over 2\tau} (T-
\tilde  T)+
                         g \hat L_{\alpha}R_{\alpha}]dx
\eeq
where $g$ is a `coupling constant',
\beq
          g={\pi\over 2\surd(k\bar k)} {(1-\tau^2)\over \tau}.
\eeq
(We can trade $\tau$ for  $g$, so that our theory is
parameterized
 by $g,k,\bar k$). Although the first two terms in the
hamiltonian density
are finite for states constructed from $|0>$, the last term is
divergent.
This is  because,   the operator
 $\int \hat L\cdot \hat R dx=\sum_m \hat L_{m\alpha} \hat R_{-
m\alpha}$
consists of two creation operators when $m<0$;
the correction to the energy  levels in  perturbation theory
is divergent at order $g^2$.
 One must then
introduce a cut--off
$\Lambda$ and define a regularized hamiltonian
\beq
     H_{\Lambda}=\int [Z(g(\Lambda),\Lambda) (T-\tilde  T)+
                         g(\Lambda) \hat
L_{\alpha}R_{\alpha}]dx.
\eeq
The coupling constant and the `wave--function' renormalization
$Z$ must
 depend on $\Lambda$ in such a way that the energy levels
remain finite
 as $\Lambda\to \infty$. This procedure will break scale
invariance.

A particular way to construct  representations of
the current algebra
is to use fermions. Let us
      introduce fermions transforming under the
representations
$r,\tilde r$ of $\un{G}$:
\beq
     \lb b_{m}^{i},b_{n}^{j}]_{+}=\delta^{ij}\delta_{m,-n}
\quad \lb \tilde  b_{m}^{\tilde  i},\tilde  b_{n}^{\tilde
j}]_{+}=
     \delta^{\tilde  i\tilde  j}\delta_{m,-n}.
\eeq
The corresponding field operators $\psi, \tilde  \psi$ are
defined by
\beq
     \psi_i(x)=\sum_m b_m^ie_m^{\half}(x),\quad
  \tilde \psi_{\tilde i}(x)=\sum_m \tilde b_m^{\tilde i}e_m^{\half}(x)
\eeq
and satisfy the Canonical Anti--commutation Relations (CAR).
We will assume that the representations satisfy the GNO
condition
\cite{god}. That is, $r,\tilde r$ are isomorphic to the
representation on  the tangent
space of a symmetric space of $G$. The most obvious choice is
that
$r$ ($\tilde  r$)  is the direct sum of $k$($\bar k$) copies of
the
fundamental representation.

We find a representation of the CAR satisfying
\beq
     b_{m}^i|0>=0\;\;{m>0},\quad \tilde  b_{m}^{\tilde
i}|0>=0\;\; m<0.
\eeq
Then we define  normal ordering by
\beqs
  \cirs b_{m}^{i}b_{n}^{j}\cirs &=&-b_{n}^{j}b_{m}^{i}
\;\;\mbox{if $m>0$ and
          $n>0$} \nn \\
                &=&\half [b_{0i},b_{0j}] \;\;\mbox{if}
                                   \;m=n=0\\
                       &=&b_{m}^{i}b_{n}^{j} \mbox{
otherwise}\nonumber
\eeqs
 and
\beqs
  \cirs \tilde  b_{m}^{\tilde  i}\tilde  b_{n}^{\tilde  j}\cirs &=&-\tilde
b_{n}^{\tilde  j}
          \tilde  b_{m}^{\tilde  i} \;\;\mbox{if $m<0$ and
          $n>0$} \nn \\
                &=&\half [\tilde  b_{0\tilde  i},\tilde  b_{0\tilde  j}]
\;\;\mbox{if}
                                   \;m=n=0\\
                       &=&\tilde  b_{m}^{\tilde  i}\tilde  b_{n}^{\tilde  j}
\mbox{     otherwise}\nonumber
\eeqs

Then, we can construct a
representation of the current algebra:
\beq
     \hat L_{\alpha}={i\over 2}\cirs \psi^i
r_{ij\alpha}\psi^j\cirs
\quad \hat R_{\alpha}={i\over 2}\cirs
          \tilde  \psi^{\tilde  i} \tilde  r_{\tilde  i\tilde
j\alpha}
\tilde  \psi^{\tilde  j}\cirs
\eeq
This  representation is finitely reducible if $r$ and $\tilde
r$ are isomorphic
to the representations of $\un{G}$ on the tangent space of
symmetric spaces
(the GNO condition). The constants $k,\bar k$ are given by
the
Dynkin index (quadratic Casimirs) of the representations
$r,\tilde  r$.
 They are, of course, positive integers.
 When the GNO condition is satisfied, moreover, the quantities
$\hat T$ and
$\hat {\tilde  T}$ form the stress tensor of a free fermion field
\cite{god}.
\beq
     \hat T= \cirs \half {\pdr \psi\over \pdr x}\psi\cirs\quad
\hat T= \cirs \half {\pdr \tilde  \psi\over \pdr x}\tilde
\psi\cirs
\eeq
Thus the hamiltonian becomes
\beqs
     H&=& Z(g)\int \biggl\{\cirs\half\dd{\psi}{x}\psi\cirs-
\cirs\half\dd{\tilde {\psi}}{x}\tilde {\psi}\cirs\biggr\}dx
                                             \nn \\
     & &-\frac{g}{4}\int \cirs
\psi_{i}\rho_{ij\alpha}\psi_{j}(x)\cirs
           \cirs \tilde {\psi}_{k}
\rho_{kl\alpha}\tilde {\psi}_{l}(x)\cirs dx.
                                        \label{th}
\eeqs
This  is the hamiltonian of a  non--abelian Thirring model. The
first term is the free hamiltonian except for a `wave--function'
renormalization $Z(g)$. The second is the current--current
coupling of the
Thirring model. With this identification, $g$ is
the Thirring
model coupling  constant. We have to use regularization and
renormalization
 to get a well--defined theory.
 The Thirring model is  usually defined with left and right
sectors in
the same representation. Our model with $k\neq \bar k$ will
have different representations
 and hence will violate parity in general.

Since the theory is not expected to be finite in general, we expect conformal
invariance to be broken in the quantum theory. The usual Thirring model has an
ultraviolet stable fixed point, so we should expect to have such a fixed point
in this theory as well.

We briefly comment on scale invariance of the quantum theory we are
describing. We don't find new fixed points for the $\beta$ function, but
we expect to recover the infrared (IR) and the ultraviolet (UV) fixed points
found in the conventional formalism, respectively in the limits
$$
\tau \longrightarrow 0,~~~~~~~~\rho=\pm 1~~~~~(IR)
$$
$$
\tau \longrightarrow \infty,~~~~~~~~\rho=\pm 1~~~~~(UV)
$$

In these limits $\lambda^2$ tends to
\beqs
\lambda^2 \longrightarrow  \frac{2 \pi}{k - \bar k} ~~&\tau& \rightarrow 0
\nn \\
0~~~&\tau& \rightarrow \infty.
\eeqs
However for $\tau \rightarrow 0, \infty$, our transformations
are singular, so that our argument is not rigorously proven.

We mention also a quantization based on a non--unitary representation of the
current algebra.  If we pick highest weight representations for both $L$ and
$R$, we will have a pseudo--unitary representation, which is unitary with
respect to a norm that is not positive. Although such a representation will
have negative norm states, we mention this possibility because of a curious
realization of conformal invariance ( which we will call "half--conformal"
invariance). Recall that the values $\rho=\pm 1 $  correspond to  conformally
invariant  theories in the conventional formalism. In the more general unitary
quantization we described earlier, for a generic value of $\tau$, even these
points do not describe conformal
field theories. If $\rho=-1$, $\Theta$ and $\tilde \Theta$ are given by
\beq
\Theta=-\frac{\lambda^2}{2} (1-\tau^2)^2 (L+R)^2,~~~~
\tilde \Theta=-\frac{\lambda^2}{2}  [(1+\tau)^2 L+(1-\tau)^2 R]^2
\eeq
but, unlike the case of highest and lowest weight representations,
if $L$ and $R$ both  form a highest weight representation, the operator
$\Theta$ can now be defined without any divergence, by normal ordering.
It will also satisfy the Virasoro algebra,  with the usual rescaling of the
constant factor. However, even when $\rho=-1$, $\tilde \Theta$ does not satisfy
the Virasoro algebra, so we still do not have full conformal invariance.
If $\rho=+1$, instead, $\tilde \Theta$ will satisfy the Virasoro algebra, while
$\Theta$ will not. Thus  in this version of the theory, we get half conformal
invariance when $\lambda$ satisfies
\beq
     {2\pi  \over \lambda^2}=|k-\bar k|
\frac {16 k \bar k} {(\surd k+\surd {\bar k})^4}.
\eeq

By simple arguments it is conjectured that, if the theory is also Lorentz
invariant, this is all what we need for scale invariance. Namely
`half--conformal invariance' and Lorentz invariance guarantee the theory to
be scale invariant. Usually, unitary scale invariance theories are also
fully conformal invariant \cite{Polchinski}. Our example does not contradict
that expectation, since the half--conformal invariant theory is not unitary.

\section{Remarks}
We wish to make a comment about the algebra \eqn{LL}--\eqn{LR}. We have so far
assumed that the parameter $\tau$ is real. It is an interesting
fact that even when the field variable $g$ is valued in a compact Lie group,
as in our case,
this is in fact not necessary. In the current algebra above, $\tau$ can be
purely imaginary; only $\tau^2$ appearing in the current algebra. The
parameters
$a,\eps,\mu$ also remain real when we allow $\tau$ to be purely imaginary. In
this case the algebra $\un{C_2}$ is not isomorphic  to a sum of two affine Lie
algebras based on $\un{G}$; instead it  becomes isomorphic to the affine Lie
algebra of the complexification of  $\un{G}$ ( but still viewed as a real
Lie algebra). Thus if $G=SU(2)$, we get the affine Lie algebra of $SL_2(C)$ or
$O(1,3)$. The parameter $k$ becomes complex and $\bar k=-k^*$, where $k^*$ is
the complex conjugate.
Thus we find the remarkable result that a WZW  model with compact target space
can have a non--compact current algebra. Representations of such a current
algebra are  typically non--unitary, \cite{lykken,gawedsky} so that
quantization along these lines may not be interesting (there are also unitary
representations of zero central charge\cite{jacobsen}, but they are not useful
for our purpose).
On the other hand, such current algebras seem to be  a good starting point for
quantization in the sense of Drinfeld \cite{reshetikin}; this possibility is
being studied.

\bigskip

\noindent {\em Aknowledgments} P. V. whishes to thank the
Department of Physics of the University of Rochester for hospitality.
S.G.R.  thanks the mathematical physics group at the University of Naples
for stimulation and hospitality. \DOE

\end{document}